\documentclass[conference]{IEEEtran}
\IEEEoverridecommandlockouts
\usepackage{cite}
\usepackage{algorithmic}
\usepackage{graphicx}
\usepackage{textcomp}
\usepackage{xcolor}
\def\BibTeX{{\rm B\kern-.05em{\sc i\kern-.025em b}\kern-.08em
    T\kern-.1667em\lower.7ex\hbox{E}\kern-.125emX}}
\graphicspath{{figures/}}
\usepackage{subcaption}
\usepackage{comment}

\usepackage{amsmath,amsfonts,stmaryrd,dsfont}
\usepackage{multirow}

\usepackage[linesnumbered,algoruled]{algorithm2e}

\usepackage{array}

\usepackage{url,cite}
\usepackage[colorlinks=true, allcolors=blue]{hyperref}

\begin{document}

\title{Spectrogram Inversion for Audio Source Separation via Consistency, Mixing, and Magnitude Constraints}

\author{\IEEEauthorblockN{Paul Magron}
\IEEEauthorblockA{\textit{Université de Lorraine, CNRS, Inria, LORIA} \\
Nancy, France \\
paul.magron@inria.fr}
\and
\IEEEauthorblockN{Tuomas Virtanen}
\IEEEauthorblockA{\textit{Audio Research Group, Tampere University} \\
Tampere, Finland \\
tuomas.virtanen@tuni.fi}
}

\maketitle

\begin{abstract}
Audio source separation is often achieved by estimating the magnitude spectrogram of each source, and then applying a phase recovery (or \emph{spectrogram inversion}) algorithm to retrieve time-domain signals. Typically, spectrogram inversion is treated as an optimization problem involving one or several terms in order to promote estimates that comply with a consistency property, a mixing constraint, and/or a target magnitude objective. Nonetheless, it is still unclear which set of constraints and problem formulation is the most appropriate in practice. In this paper, we design a general framework for deriving spectrogram inversion algorithm, which is based on formulating optimization problems by combining these objectives either as soft penalties or hard constraints. We solve these by means of algorithms that perform alternating projections on the subsets corresponding to each objective/constraint. Our framework encompasses existing techniques from the literature as well as novel algorithms. We investigate the potential of these approaches for a speech enhancement task. In particular, one of our novel algorithms outperforms other approaches in a realistic setting where the magnitudes are estimated beforehand using a neural network.
\end{abstract}


\IEEEpeerreviewmaketitle

\section{Introduction}

Audio source separation consists in extracting the underlying \textit{sources} that add up to form an observed audio \textit{mixture}. Typical source separation approaches use a deep neural network (DNN) to estimate a nonnegative mask that is applied to a time-frequency (TF) representation of the audio mixture, such as the short-time Fourier transform (STFT)~\cite{Wang2018c}. Alternatively, complex-valued DNNs jointly process the real and imaginary parts of the STFT~\cite{Hu2020}, and end-to-end networks operate in the time domain directly, where the STFT is replaced with learned filterbanks~\cite{Luo2019,Pariente2020}. Nonetheless, it was recently shown that nonnegative TF masking remains interesting, since it yields competitive results with lighter and more interpretable networks~\cite{Heitkaemper2020,Cord2022}.

Such a masking results in assigning the mixture's STFT phase to each isolated source, which induces residual interference and artifacts in the estimates. Consequently, a significant research effort has been put on phase recovery, also called \emph{spectrogram inversion}. While recent approaches mostly rely on deep phase models~\cite{Masuyama2023} or neural vocoders~\cite{Kong2020}, in this work we focus on optimization-based iterative algorithms~\cite{Gunawan2010}. Indeed, these remain powerful since they can be either used as a light post-processing, unfolded within end-to-end systems~\cite{Wang2018,Masuyama2021}, or combined with the above-mentioned deep phase models.

Iterative spectrogram inversion algorithms are usually derived as solutions to optimization problems involving one or several terms that promote desirable properties in the TF domain. For instance, the multiple input spectrogram inversion (MISI) algorithm~\cite{Gunawan2010} minimizes a measure of magnitude spectrogram mismatch under a mixing constraint (the estimates must add up to the mixture). Alternatively, the authors in~\cite{Wang2019} relax the mixing constraint into a soft penalty that is added to a \emph{consistency}~\cite{LeRoux2010a} term. Conversely,~\cite{Magron2018} considers a hard magnitude constraint and discards the consistency criterion.  Nonetheless, it is still unclear which set of constraints and problem formulation is the most appropriate in practice.

In this paper, we design a general framework for deriving spectrogram inversion algorithms. We consider three objectives in the TF domain: \emph{mixing}, \emph{consistency}, and \emph{magnitude match}, and we formulate optimization problems by combining these objectives either as soft penalties or hard constraints. We then derive an auxiliary function for each term, which allows to solve the corresponding optimization problems by means of alternating projection algorithms. While this framework encompasses existing techniques from the literature, it also allows to derive novel algorithms. We experimentally assess the potential of these algorithms for a speech enhancement task on an freely available audio corpus. In particular, one of our novel algorithms outperforms baseline MISI variants~\cite{Gunawan2010,Wang2019} in a realistic setting where the magnitudes are estimated beforehand using a DNN.

The rest of this paper is structured as follows. Section~\ref{sec:framework} introduces the proposed framework, and algorithms are derived in Section~\ref{sec:algos}. Section~\ref{sec:exp} presents the experimental results. Finally, Section~\ref{sec:conclu} concludes the paper.

\section{Proposed framework}
\label{sec:framework}

\subsection{Problem setting}

Let us consider a monaural instantaneous mixture model:
\begin{equation}
\mathbf{X} = \sum_{j=1}^J \mathbf{S}_j,
\label{eq:mixmodel}
\end{equation}
where $\mathbf{X} \in \mathbb{C}^{F \times T}$ is the mixture STFT, and ${\mathbf{S}_j \in \mathbb{C}^{F \times T}}$ are the $J$ sources matrices, whose entries are denoted $x_{f,t}$ and $s_{j,f,t}$. $F$ and $T$ respectively denote the number of frequency channels and time frames of the STFT. We assume that the sources' STFT magnitudes $\mathbf{V}_j$ have been estimated beforehand, e.g., via a DNN. Then, spectrogram inversion consists in estimating the complex-valued sources $\mathbf{S} =  \{\mathbf{S}_1,\ldots,\mathbf{S}_J \}$ in order to further invert these for retrieving time-domain signals. To perform this task, we search for source estimates that comply with the following properties:

\begin{itemize}
\item \textbf{Mixing}: the estimates should sum up to the mixture according to~\eqref{eq:mixmodel}, so that there is no creation or destruction of energy overall. Such estimates are said to be \emph{conservative}.
\item \textbf{Consistency}: the estimates should be \emph{consistent}~\cite{LeRoux2010a}, that is, the corresponding complex-valued matrices should be the STFT of time-domain signals.
\item \textbf{Magnitude match}: the magnitudes of the estimates should remain close to the target magnitudes $\mathbf{V}_j$ that have been estimated beforehand. 
\end{itemize}

We propose to define a loss function corresponding to each objective in an optimization framework, and to combine these either as soft penalties or hard constraints. To solve the corresponding optimization problems, we resort to the auxiliary function method, which has shown powerful for spectrogram inversion~\cite{Magron2020, Magron2018, LeRoux2010a}. In a nutshell, if we consider minimization of a function $\phi$ with parameters $\theta$, this approach consists in constructing and minimizing an \emph{auxiliary function} $\phi^+$ with additional \emph{auxiliary} parameters $\tilde{\theta}$ such that $\forall \theta, \quad \phi(\theta) = \min_{\tilde{\theta}} \phi^+(\theta, \tilde{\theta})$. Then, it can easily be shown that $\phi$ is non-increasing under the following update scheme:
\begin{equation}
   \tilde{\theta} \leftarrow \arg \min_{\tilde{\theta}} \phi^+(\theta, \tilde{\theta}) \quad \text{and} \quad \theta \leftarrow \arg \min_{\theta} \phi^+(\theta, \tilde{\theta}).
\end{equation}
For a clarity purpose, this paper focuses on formulating the optimization problems and providing the algorithms' updates; a supporting document details the mathematical derivations.\footnote{\url{https://magronp.github.io/files/2023specinv_sup.pdf}}

\subsection{Mixing error}
\label{sec:framework_mix}

We consider the following mixing error:
\begin{equation}
h(\mathbf{S}) =  ||\mathbf{X} - \sum_j \mathbf{S}_j||^2,
\label{eq:mix_error}
\end{equation}
where $||.||$ is the Frobenius norm. Let us consider auxiliary parameters $\mathbf{Y} = \{\mathbf{Y}_1, \ldots, \mathbf{Y}_J \}$ such that $\sum_j \mathbf{Y}_j = \mathbf{X}$, and positive weights $\mathbf{\Lambda}_j = \{ \lambda_{j,f,t} \}_{f,t}$ such that $\sum_j \lambda_{j,f,t} = 1$. We define $h^+$ as follows:
\begin{equation}
h^+(\mathbf{S},\mathbf{Y}) = \sum_{j,f,t} \frac{|y_{j,f,t}-s_{j,f,t}|^2}{\lambda_{j,f,t}}.
\end{equation}
Then, using the Jensen inequality, one can show that~\cite{Magron2018}:
\begin{equation}
h(\mathbf{S}) = \min_{\mathbf{Y}} h^+(\mathbf{S},\mathbf{Y}) \quad \text{s. t.} \quad \sum_j \mathbf{Y}_j = \mathbf{X},
\end{equation}
which shows that $h^+$ is an auxiliary function for $h$. Besides, the auxiliary parameters' update is given by~\cite{Magron2018}:
\begin{equation}
\mathbf{Y}_j = \mathbf{S}_j + \mathbf{\Lambda}_j \odot (\mathbf{X} - \sum_k \mathbf{S}_k),
\label{eq:update_Y}
\end{equation}
where $\odot$ denotes the element-wise matrix multiplication.

\subsection{Inconsistency}
\label{sec:framework_incons}

To promote consistent estimates, we consider the inconsistency measure defined in~\cite{LeRoux2010a}: $i(\mathbf{S}) = \sum_j ||\mathbf{S}_j - \mathcal{G}(\mathbf{S}_j) ||^2$, where $\mathcal{G} = \text{STFT} \circ \text{iSTFT}$. It is proven~\cite{LeRoux2010a} that $\mathcal{G}(\mathbf{S}_j)$ is the closest consistent matrix to $\mathbf{S}_j$ in a least-square sense, that is:
\begin{equation}
||\mathbf{S}_j - \mathcal{G}(\mathbf{S}_j) ||^2 = \min_{\mathbf{Z}_j} ||\mathbf{S}_j - \mathbf{Z}_j||^2 \quad \text{s. t.} \quad \mathbf{Z}_j \in \mathcal{I},
\end{equation}
where $\mathcal{I}$ is the image set of the STFT operator. Therefore, $i^+(\mathbf{S}, \mathbf{Z}) = \sum_j ||\mathbf{S}_j - \mathbf{Z}_j||^2$ is an auxiliary function for $i$, and the auxiliary parameters' update is given by:
\begin{equation}
\mathbf{Z}_j = \mathcal{G}(\mathbf{S}_j).
\label{eq:update_Z}
\end{equation}

\subsection{Magnitude mismatch}

Finally, we consider the following loss for characterizing the magnitude mismatch:\footnote{Note that we recently investigated alternative magnitude discrepancy measures~\cite{Magron2021}, but we focus on the Frobenius norm in this study.}
\begin{equation}
m(\mathbf{S}) = \sum_{j} || |\mathbf{S}_{j}| - \mathbf{V}_{j} ||^2.
\label{eq:mag_function}
\end{equation}
We introduce a set of auxiliary parameters $\mathbf{U}_j$ such that $|\mathbf{U}_j|=\mathbf{V}_j$. Then, drawing on~\cite{Magron2020}, we have
\begin{equation}
|| |\mathbf{S}_{j}| - \mathbf{V}_{j} ||^2 = \min_{\mathbf{U}_j} ||\mathbf{S}_j - \mathbf{U}_j||^2 \quad \text{s. t.} \quad |\mathbf{U}_j| = \mathbf{V}_j,
\end{equation}
and the minimum is reached for:
\begin{equation}
\mathbf{U}_j = \frac{\mathbf{S}_j}{|\mathbf{S}_j|} \odot \mathbf{V}_j.
\label{eq:update_U}
\end{equation}
This proves that $m^+(\mathbf{S},\mathbf{U}) = \sum_{j} || \mathbf{S}_j - \mathbf{U}_j ||^2$ is an auxiliary function for $m$.

\section{Algorithms derivation}
\label{sec:algos}

\subsection{Mixing and consistency as objectives}
\label{sec:mix+incons}

First, let us ignore the magnitude constraint, and consider the following problem:
\begin{equation}
\min_{\mathbf{S}}  h(\mathbf{S}) + \sigma i(\mathbf{S}),
\label{eq:cost_smooth_nomag}
\end{equation}
where $\sigma \geq 0 $ is a weight adjusting the relative importance of the consistency constraint (for notation purposes, $\sigma = + \infty$ corresponds to an inconsistency objective only). Using our proposed framework, the problem rewrites:
\begin{equation}
\min_{\mathbf{S},\mathbf{Y},\mathbf{Z}} h^+(\mathbf{S}, \mathbf{Y}) + \sigma  i^+(\mathbf{S}, \mathbf{Z}) \text{ s. t. } \sum_j \mathbf{Y}_j = \mathbf{X} \text{ and } \mathbf{Z}_j \in \mathcal{I}.
\label{eq:cost_smooth_nomag_aux}
\end{equation}
The updates for $\mathbf{Y}$ and $\mathbf{Z}$ have been derived previously (see Section~\ref{sec:framework_mix} and~\ref{sec:framework_incons}), thus we only need to obtain the update for $\mathbf{S}$. To do so, we set the partial derivative of the objective function in~\eqref{eq:cost_smooth_nomag_aux} with respect to $\mathbf{S}_j$ at $0$ and solve, which yields:
\begin{equation}
\mathbf{S}_j = \frac{\mathbf{Y}_j + \sigma \mathbf{\Lambda}_j \odot \mathbf{Z}_j}{1 + \sigma \mathbf{\Lambda}_j},
\label{eq:update_mix_cons_nomag}
\end{equation}
where division is meant element-wise. Therefore, alternating updates~\eqref{eq:update_Y},~\eqref{eq:update_Z}, and~\eqref{eq:update_mix_cons_nomag} yields an iterative procedure that solves~\eqref{eq:cost_smooth_nomag}. We call it \texttt{Mix+Incons}, and we remark that:
\begin{itemize}
    \item This procedure is similar to the first version of the consistent Wiener filtering~\cite{LeRoux2010a}, which forces the estimates to be close to Wiener filter estimates instead of $\mathbf{Y}_j$. Nonetheless, both sets of estimates are conservative.
    \item Choosing $\sigma=0$ or $\sigma=+ \infty$ leads to $\mathbf{S}_j = \mathbf{Y}_j$ and $\mathbf{S}_j = \mathbf{Z}_j$, respectively. These non-iterative estimators are termed ``mixture-consistent projection" and ``STFT-consistent projection" in~\cite{Wisdom2019}. Thus, the general update given by~\eqref{eq:update_mix_cons_nomag} allows for a smooth trade-off between these.
\end{itemize}

\subsection{Mixing and consistency with a hard magnitude constraint}
\label{sec:mix+incons_hardmag}

We now incorporate an additional hard magnitude constraint into~\eqref{eq:cost_smooth_nomag} by means of the method of Lagrange multipliers. This results in finding a critical point for:
\begin{equation}
h^+(\mathbf{S}, \mathbf{Y}) + \sigma  i^+(\mathbf{S}, \mathbf{Z}) + \sum_{j,f,t} \delta_{j,f,t} (|s_{j,f,t}|^2-v_{j,f,t}^2),
\label{eq:mix_incons_hardmag}
\end{equation}
where $\{ \delta_{j,f,t} \}_{j,f,t}$ are the Lagrange multipliers. We set the partial derivative of~\eqref{eq:mix_incons_hardmag} with respect to $\mathbf{S}_j$ at $0$ and solve, which yields:
\begin{equation}
\mathbf{S}_j = \frac{\mathbf{Y}_j + \sigma \mathbf{\Lambda}_j \odot \mathbf{Z}_j}{|\mathbf{Y}_j + \sigma \mathbf{\Lambda}_j \odot \mathbf{Z}_j|} \odot \mathbf{V}_j.
\end{equation}
We call this procedure \texttt{Mix+Incons\_hardMag}. Note that:
\begin{itemize}
\item  This procedure is equivalent to the modified MISI algorithm from~\cite{Wang2019}, which however treats the weights $\mathbf{\Lambda}$ as unknown parameters and updates them at each iteration. 
\item If $\sigma=+ \infty$, the procedure boils down to applying the well-known Griffin-Lim update~\cite{Griffin1984} to each source independently without mixing constraint.
\item If $\sigma=0$, the procedure reduces to our previous ``PU-Iter"~\cite{Magron2018}, which discards the consistency constraint.
\end{itemize}

\subsection{Consistency objective with a hard mixing constraint}
\label{sec:inc_mixhard}

Now, let us consider an inconsistency-only objective function, where mixing is treated as a hard constraint. Note that in this setup, we do not consider an additional hard magnitude constraint since this yields an ill-posed problem.\footnote{Indeed, one can verify on a simple example ($J=2$, $v_1=v_2=1$, and $x=4$) that there is no solution that satisfies both constraints in general.} As above, we treat this problem with the method of Lagrange multipliers, which eventually yields:
\begin{equation}
\mathbf{S}_j = \mathbf{Z}_j + \frac{1}{J} (\mathbf{X} - \sum_k \mathbf{Z}_k),
\label{eq:cons_hardmix_up}
\end{equation}
where the update for $\mathbf{Z}$ is given by~\eqref{eq:update_Z}. We call this method \texttt{Incons\_hardMix}, and we remark that:
\begin{itemize}
    \item This approach is non-iterative, since the set of consistent matrices is a vector space and $\mathbf{Z}_j$ is consistent by construction.
    \item It is equivalent to the successive application of the STFT- and mixture-consistent projections used in~\cite{Wisdom2019}.
    \item The update~\eqref{eq:cons_hardmix_up} is similar to~\eqref{eq:update_Y} with fixed weights ${\mathbf{\Lambda}_j = 1/J}$, which is expected when using mixing as a hard constraint~\cite{Gunawan2010}.
\end{itemize}

\subsection{Magnitude objective with a hard mixing constraint}

Finally, we consider the magnitude mismatch as the main objective under a hard mixing constraint. Since incorporating an additional hard consistency constraint would eventually yield MISI~\cite{Magron2020}, we focus here on a soft consistency penalty:
\begin{equation}
\min_{\mathbf{S}} m(\mathbf{S}) + \sigma i(\mathbf{S}) \quad \text{s. t.} \quad \sum_j \mathbf{S}_j = \mathbf{X}.
\end{equation}
Still using the method of Lagrange multipliers, we obtain:
\begin{align}
\mathbf{W}_j &= \frac{1}{1+\sigma} \left (\mathbf{U}_j + \sigma \mathbf{Z}_j \right),\\
\mathbf{S}_j &= \mathbf{W}_j + \frac{1}{J} \left( \mathbf{X} - \sum_k \mathbf{W}_k \right),
\end{align}
where $\mathbf{Z}_j$ and $\mathbf{U}_j$ are given by~\eqref{eq:update_Z} and~\eqref{eq:update_U}. We call this procedure \texttt{Mag+Incons\_hardMix}.

\noindent \emph{Remark}: Let us note that if we discard the consistency constraint ($\sigma=0$) and initialize the estimates using an amplitude mask (see Section~\ref{sec:exp_protocol}), then the estimator becomes:
\begin{equation}
\mathbf{S}_j = \left( \mathbf{V}_j + \frac{1}{J}(|\mathbf{X}|-\sum_k \mathbf{V}_k) \right) \frac{\mathbf{X}}{|\mathbf{X}|},
\label{eq:update_s_closed}
\end{equation}
which is non-iterative and assigns the mixture's phase to each source, therefore it does not improve phase recovery.

\subsection{Summary of the algorithms}

\begin{table*}[ht!]
	\center
    \caption{Alternating projection algorithms for spectrogram inversion, where particular cases from the literature are indicated in \textit{italics}. $\mathbf{\Lambda}_j$ denotes mixing weights that can be hand-tuned. Note that all algorithms exhibit a similar computational cost, which is dominated by the calculation of the STFT / iSTFT, thus consistency-free approaches are slightly faster.}
	\label{tab:algos}
	\begin{tabular}{lccccc}
    \hline
    \hline
		Algorithm & Reference & Consistency weight & Mixing weights & Iterative & Update formula \\
		\hline
\\[-0.5em]
\texttt{MISI}          & \cite{Gunawan2010}      & no & $1/J$                & yes & $\displaystyle\mathcal{P}_{\text{mix}}(\mathcal{P}_{\text{mag}}(\mathcal{P}_{\text{cons}}(\mathbf{S})))$ \\
\\[-1em]
\texttt{Mix+Incons} & & $\sigma$ & $\{ \mathbf{\Lambda}_j \}_j$ & yes & $\displaystyle \frac{1}{1+\sigma \mathbf{\Lambda}} \odot (\mathcal{P}_{\text{mix}}(\mathbf{S}) + \sigma \mathbf{\Lambda} \odot \mathcal{P}_{\text{cons}}(\mathbf{S}))$ \\
\\[-1em]
\hspace{0.7em} \textit{Mixture-consistent projection} & \cite{Wisdom2019}  & $\sigma = 0$ &  & no & $\displaystyle\mathcal{P}_{\text{mix}}(\mathbf{S})$\\
\\[-0.5em]
\hspace{0.7em} \textit{STFT-consistent projection} & \cite{Wisdom2019} & $\sigma = +\infty$ & & no & $\displaystyle\mathcal{P}_{\text{cons}}(\mathbf{S})$ \\
\\[-0.5em]
\texttt{Mix+Incons\_hardMag}     &   & $\sigma$ & $\{ \mathbf{\Lambda}_j \}_j$ & yes & $\displaystyle\mathcal{P}_{\text{mag}}(\mathcal{P}_{\text{mix}}(\mathbf{S}) + \sigma \mathbf{\Lambda} \odot \mathcal{P}_{\text{cons}}(\mathbf{S}))$ \\
\\[-0.5em]
\hspace{0.7em} \textit{Modified MISI}     & \cite{Wang2019} & $\sigma$ & optimized & yes & $\displaystyle\mathcal{P}_{\text{mag}}(\mathcal{P}_{\text{mix}}(\mathbf{S}) + \sigma \mathbf{\Lambda} \odot \mathcal{P}_{\text{cons}}(\mathbf{S}))$ \\
\\[-0.5em]
\hspace{0.7em} \textit{PU-Iter}    &  \cite{Magron2018}    & $\sigma=0$ & $\{ \mathbf{\Lambda}_j \}_j$ & yes & $\displaystyle\mathcal{P}_{\text{mag}}(\mathcal{P}_{\text{mix}}(\mathbf{S}))$ \\
\\[-0.5em]
\texttt{Incons\_hardMix}     &  \cite{Wisdom2019} & no & $1/J$ or $\{ \mathbf{\Lambda}_j \}_j$ or learned   & no & $\displaystyle\mathcal{P}_{\text{mix}}(\mathcal{P}_{\text{cons}}(\mathbf{S}))$ \\
\\[-1em]
\texttt{Mag+Incons\_hardMix}     &     & $\sigma$ & $1/J$                & yes & $\displaystyle \mathcal{P}_{\text{mix}} \left( \frac{1}{1+\sigma} (\mathcal{P}_{\text{mag}}(\mathbf{S}) + \sigma \mathcal{P}_{\text{cons}}(\mathbf{S})) \right)$ \\
\\[-1em]
       \hline
       \hline
	\end{tabular}
\end{table*}

We summarize in Table~\ref{tab:algos} the updates performed by these various algorithms using the following magnitude, consistency, and mixing projectors:
\begin{align}
&\mathcal{P}_{\text{mag}}(\mathbf{S}) = \left\lbrace \frac{\mathbf{S}_j }{|\mathbf{S}_j |} \odot \mathbf{V}_j \right\rbrace_j \\
&\mathcal{P}_{\text{cons}}(\mathbf{S}) = \left\lbrace \mathcal{G}(\mathbf{S}_j)  \right\rbrace_j \\
&\mathcal{P}_{\text{mix}}(\mathbf{S}) = \left\lbrace \mathbf{S}_j + \mathbf{\Lambda}_j \odot (\mathbf{X} - \sum_k \mathbf{S}_k)  \right\rbrace_j.
\end{align}

\section{Experiments}
\label{sec:exp}

In this section, we assess the potential of our algorithms for a speech enhancement task, a particular case of source separation with $J=2$ sources (speech and noise). Note that this framework remains applicable to other source separation scenarios such as speech~\cite{Wang2018c} or music~\cite{Stoter2019} separation.
We provide our code and sound examples online.\footnote{\url{https://github.com/magronp/spectrogram-inversion}} 

\subsection{Protocol}
\label{sec:exp_protocol}

\paragraph*{Data}
As acoustic material, we build mixtures of clean speech and noise. We randomly select $100$ utterances from the VoiceBank set~\cite{Valentini2016} to create the clean speech, and we select three real-world environments noise signals (living room, bus, and public square) from the DEMAND dataset~\cite{Thiemann2013}. For each clean speech signal, we randomly select a noise excerpt cropped at the same length than that of the speech signal. We then mix the two signals at various input signal-to-noise ratios (iSNRs) ($10$, $0$, and $-10$ dB). All audio excerpts are single-channel and sampled at $16$ kHz. The STFT is computed with a $1024$ samples-long ($64$ ms) Hann window and 75$\%$ overlap. The dataset is split into two subsets of $50$ mixtures: a \emph{validation} set, on which the consistency weight is tuned; and a \emph{test} set, on which all algorithms are evaluated.

\paragraph*{Spectrogram estimation}
 We estimate the magnitude spectrograms $\mathbf{V}_j$ with Open-Unmix~\cite{Stoter2019}, an open implementation of a three-layer BLSTM neural network, originally tailored for music source separation, and later adapted to speech enhancement. We use the pre-trained model available at~\cite{Uhlich2020} and described in~\cite{Valentini2016}. In practice, magnitudes are estimated more accurately as the iSNR increases.

\paragraph*{Methods}
All algorithms are initialized with an amplitude mask (AM), i.e., the estimated magnitudes are combined with the mixture's phase. The number of iterations is tuned on the validation set (see next section), with a maximum of $20$. For the mixing projector, any nonnegative weights $\mathbf{\Lambda}_j$ that verify the sum-to-one constraint can be used. In practice, we consider magnitude ratios $\mathbf{\Lambda}_j = \mathbf{V}_j / \sum_k \mathbf{V}_k$, since these are common in such algorithms~\cite{Magron2018, Wisdom2019} and yield the same performance as the optimized weights described in~\ref{sec:mix+incons_hardmag}.

\paragraph*{Metric} We evaluate the separation quality via the signal-to-distortion ratio (SDR) between the true clean speech $\mathbf{s}_1^\star$ and its estimate $\mathbf{s}_1$ (averaged over signals, higher is better):
\begin{equation}
    \text{SDR}(\mathbf{s}_1^\star, \mathbf{s}_1) = 20 \log_{10}\frac{\|\mathbf s_1^\star\|}{\|\mathbf s_1^\star - \mathbf s_1\|}.
\end{equation}

\subsection{Results}

\begin{figure*}
\centering
\begin{subfigure}{.99\linewidth}
\centering
\includegraphics[scale=.7]{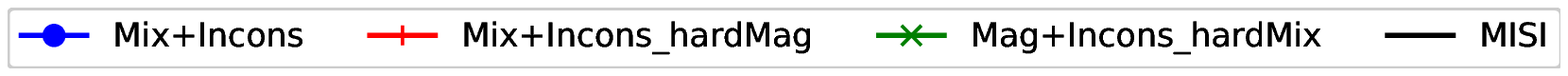}
\end{subfigure}
\begin{subfigure}{.25\textwidth}
  \centering
  \includegraphics[width=.99\linewidth]{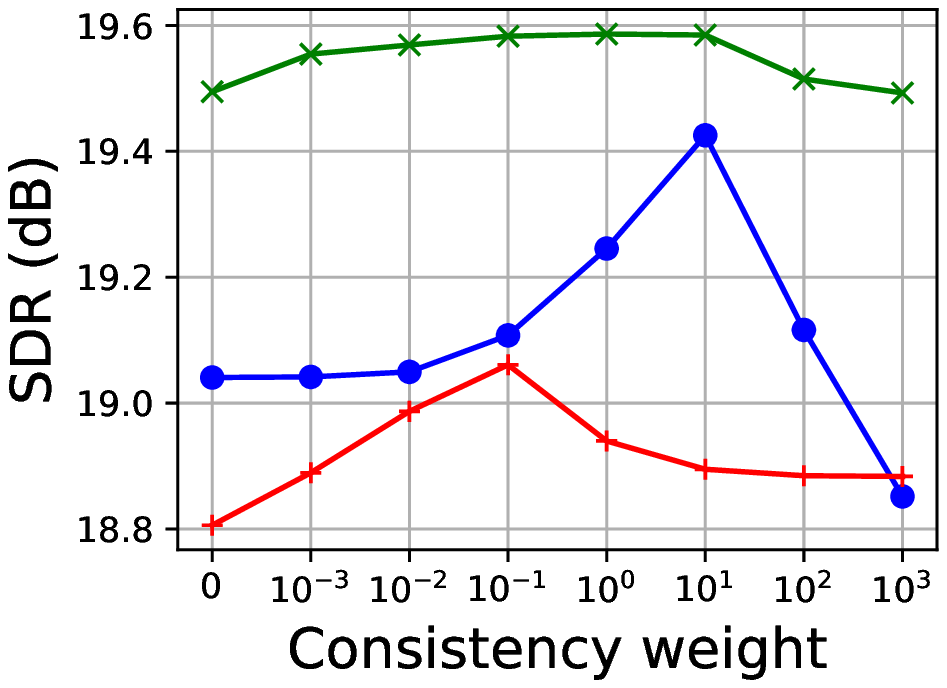}
    \caption{iSNR=10 dB}
    \label{fig:val_cons_10}
\end{subfigure}%
\begin{subfigure}{.25\textwidth}
  \centering
  \includegraphics[width=.99\linewidth]{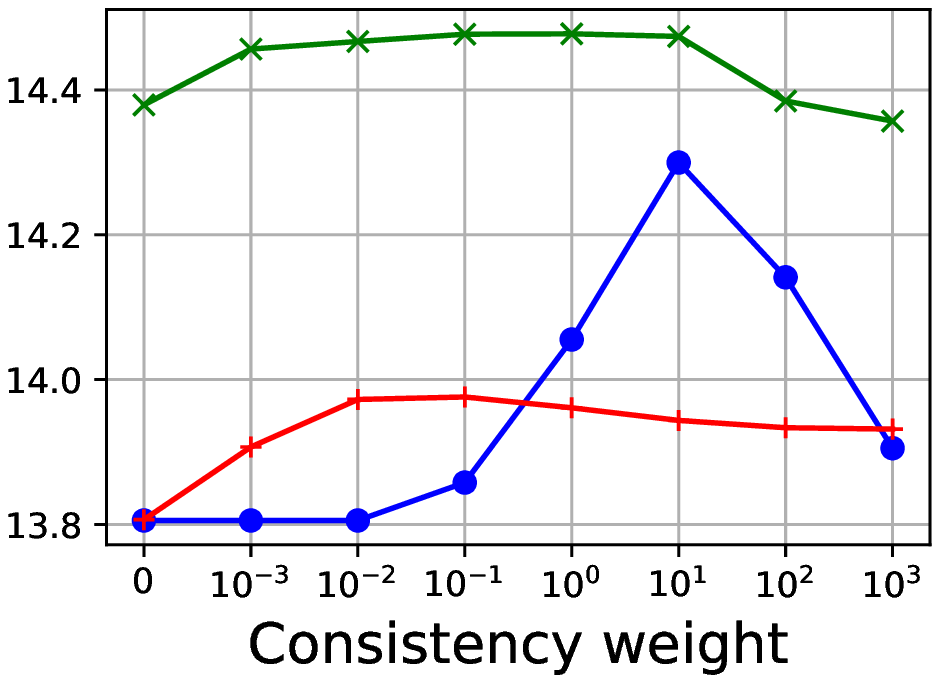}
    \caption{iSNR=0 dB}
    \label{fig:val_cons_0}
\end{subfigure}%
\begin{subfigure}{.25\textwidth}
  \centering
  \includegraphics[width=.99\linewidth]{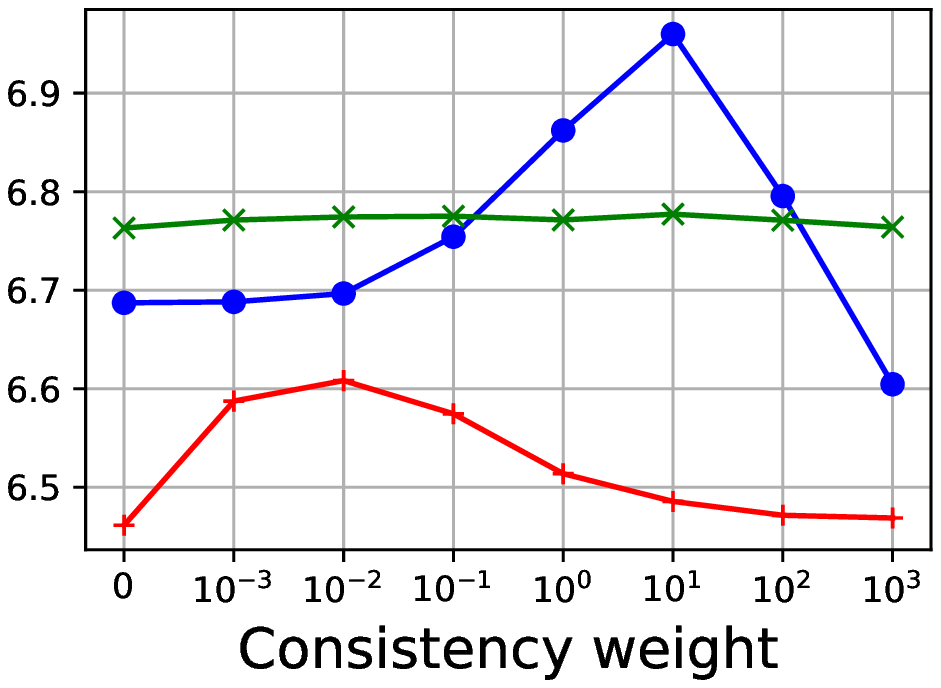}
    \caption{iSNR=-10 dB}
    \label{fig:val_cons_m10}
\end{subfigure}%
\begin{subfigure}{.25\textwidth}
  \centering
  \includegraphics[width=.99\linewidth]{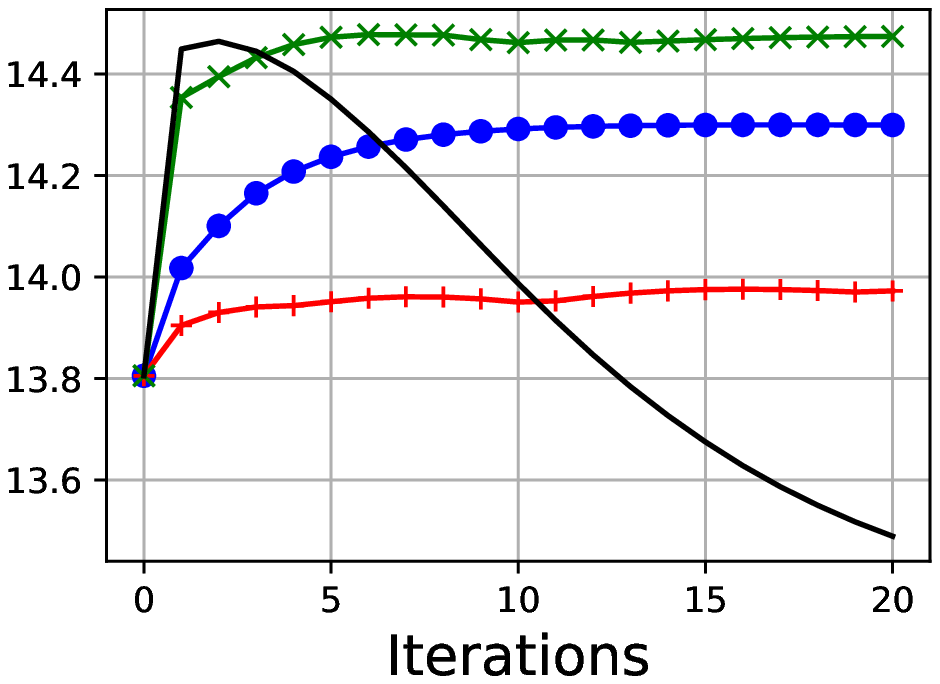}
    \caption{iSNR=0 dB}
    \label{fig:val_iter}
\end{subfigure}%
\caption{Validation SDR as a function of $\sigma$ for the optimal number of iterations at various iSNRs (\ref{fig:val_cons_10}-\ref{fig:val_cons_m10}), and validation SDR over iterations for the optimal consistency weight at iSNR=0 dB (\ref{fig:val_iter}); similar trends are observed at other iSNRs.}
\label{fig:val}
\end{figure*}

First, let us investigate the influence of the consistency weight $\sigma$ onto performance. From the results displayed in Fig.~\ref{fig:val_cons_10}-\ref{fig:val_cons_m10}, we remark that \texttt{Mag+Incons\_hardMix} exhibits a stable performance with respect to $\sigma$. Conversely, for the other algorithms, the SDR peaks at specific values of $\sigma$. In particular, a properly tuned \texttt{Mix+Incons} outperforms its special cases corresponding to $\sigma=0$ and $+\infty$, which were used in~\cite{Wisdom2019}. A similar behavior is observed for \texttt{Mix+Incons\_hardMag}, which outperforms its special case $\sigma=0$~\cite{Magron2018}. This demonstrates the interest of our framework, which allows for deriving general algorithms that outperform their special cases found in the literature.

Besides, Fig.~\ref{fig:val_iter} reveals that the algorithms exhibit a different behavior over iterations. Indeed, while for most methods, the SDR steadily increases over iterations, MISI reaches its peak performance after very few iterations, and then its performance drops severely. This was notably observed in~\cite{Wang2019}; nonetheless, the modified MISI algorithm introduced in this paper (which is similar to \texttt{Mix+Incons\_hardMag}) was compared to MISI after $200$ iterations, which is somewhat unfair to MISI. Here, we select the optimal weight and number of iterations for each algorithm in order to run them on the test set in a more fair setup.

\begin{table}[t]
	\center
    \caption{Test results (SDR in dB).}
	\label{tab:test_sdr}
	\begin{tabular}{lccc}
    \hline
    \hline
		 & iSNR$=10$ & iSNR$=0$ & iSNR$=-10$  \\
		\hline
\texttt{AM}                   & $18.7$ & $13.5$ & $7.7$ \\
\texttt{MISI}                 & $\mathbf{19.6}$ & $\mathbf{14.1}$ & $7.7$ \\
\texttt{Mix+Incons}           & $19.3$ & $13.7$ & $\mathbf{8.1}$ \\
\texttt{Mix+Incons\_hardMag}  & $18.7$ & $13.8$ & $7.9$ \\
\texttt{Incons\_hardMix}      & $\mathbf{19.6}$ & $13.9$ & $7.5$ \\
\texttt{Mag+Incons\_hardMix}  & $\mathbf{19.6}$ & $\mathbf{14.1}$ & $7.7$ \\
       \hline
       \hline
	\end{tabular}
 \vspace{-1em}
\end{table}

From the test results presented in Table~\ref{tab:test_sdr}, we remark that MISI achieves the best performance at high or moderate iSNR, i.e., when the spectrograms are rather accurately estimated. We also observe that \texttt{Mag+Incons\_hardMix} can be an interesting alternative to MISI: indeed, both algorithms perform similarly in terms of SDR, but the latter is easier to tune, as is evident from its steady behavior over iterations and stability with respect to $\sigma$ (see Fig.~\ref{fig:val}). The \texttt{Incons\_hardMix} algorithm, which was used for end-to-end source separation in~\cite{Wisdom2019}, reveals interesting at high iSNR since it is non-iterative and yields similar results to MISI. However, it performs the worst at low iSNR, where a baseline AM is preferable.  On the other hand, while \texttt{Mix+Incons\_hardMag} improves over MISI at low iSNR, it becomes less interesting when spectrograms are more accurately estimated (iSNR = $0$ or $10$ dB), which differs from the results of~\cite{Wang2019}. This difference might be explained by the usage of a different magnitude estimation technique which is of paramount importance in phase recovery~\cite{Magron2018} (spectral subtraction in~\cite{Wang2019} vs. a DNN here); and by the afore-mentioned impact of an optimized number of iterations. Finally, we note that the proposed \texttt{Mix+Incons} allows to mitigate this SDR drop at high iSNR, while it further improves the performance at low iSNR by $0.2$ dB over the previously best performing approach. It should be noted that this technique does not exploit the magnitude projector, and only relies on the estimated magnitudes $\mathbf{V}_j$ via its initialization. Therefore, the algorithm allows for deviation from these magnitude values, which might explain its good performance at low iSNR, since magnitude are estimated with a lower accuracy in this case.

\section{Conclusion}
\label{sec:conclu}

In this paper, we introduced a general framework for deriving alternating projection algorithms for spectrogram inversion using consistency, mixing, and magnitude constraints. This framework encompasses existing techniques from the literature, but also yields novel algorithms, among which some appear as promising alternatives to baseline spectrogram inversion approaches. Future work will be devoted to adapt these algorithms to operate online~\cite{Magron2020} in order to combine them with model-based phase priors~\cite{Magron2018, Masuyama2023}. We will also unfold them into deep networks for end-to-end separation~\cite{Wang2018}, where supervised learning can be leveraged to optimize the consistency and mixing weights.

\bibliographystyle{IEEEtran}
\bibliography{references}

\end{document}